\documentclass[aps,prb,preprint,groupedaddress]{revtex4-1}
\usepackage{graphicx}
\usepackage{hyperref}
\usepackage{siunitx}
\usepackage{amsmath}
\usepackage{caption}
\usepackage{subcaption}
\usepackage{natbib}
\usepackage{graphicx}
\usepackage{dcolumn}
\usepackage{bm}
\usepackage{xcolor}
\begin{document}
\setcitestyle{numbers}
\setcitestyle{square}

\title{ Electronic phase transition, vibrational properties and  structural stability of single and two polyyne chains under external electric field} 

\author{Karthik H J}
\author{Sarga P K}

\author{Swastibrata Bhattacharyya}
\email{swastibratab@goa.bits-pilani.ac.in }
\affiliation{Department of Physics, Birla Institute of Technology and Science Pilani, K. K. Birla Goa Campus, Zuarinagar, Goa, 403726, India}

\date{\today}
\begin{abstract}
\textcolor{black}{Search for one dimensional (1D) van der Waals materials has become an urgent need to meet the demand as building blocks for high performance, miniaturized, lightweight device applications.} Polyyne, a 1D atomic chain of carbon is the thinnest and strongest allotrope of carbon,  showing promising applications in new generation low dimensional devices due to the presence of a band gap. A system of two carbon chains held together by van der Waals interaction has been theoretically postulated and shows band gap tunability under structural changes \textcolor{black}{which finds applications in the realms of resistive switching and spintronics.}  In this study, we use first principles Density Functional Theory (DFT) to show a sharp semiconductor to metal transition along with the emergence of an asymmetry in the spin polarized density of states for single and two polyyne chains under a transverse electric field. The thermodynamic stability of the system has been substantiated through the utilization of \textit{Ab Initio} Molecular Dynamics (AIMD) simulations, phonon dispersion curve analyses, and formation energy calculations. Furthermore, in addition to its dynamic stability assessment, phonon calculations have served to identify Raman-active vibrational modes which offers an invaluable non-destructive experimental avenue for discerning electronic phase transitions in response to an applied electric field. Our study presents a predictive framework for the prospective utilization of one and two polyyne chains in forthcoming flexible nano-electronic and spintronic devices. \textcolor{black}{The future prospects of the system are contingent upon advancements in nano-electronics fabrication techniques and the precise construction of circuitry for harnessing spin-related applications.}
\end{abstract}

\maketitle  
\section{INTRODUCTION}
Due to the presence of three plausible hybridization states (sp$^3$, sp$^2$ and sp) of its atomic orbitals, carbon is a chemically multi-functional element. These three paradigmatic structures are diamond; graphene-like structures (graphite, graphene nanoribbons, carbon nanotubes (CNT) and fullerenes); and linear carbon chain (carbyne)\cite{heimann1999carbyne} for sp$^3$, sp$^2$ and sp  hybridization states, respectively.  Carbyne is the thinnest one-dimensional (1D) structure of carbon formed by  carbon atoms linearly arranged as a chain\cite{casari2018carbyne}.  Carbyne chain can be found in two different phases: polyyne ($\alpha$ phase) (–C$\equiv$C–), with alternate single and triple bonds, and cumulene ($\beta$ phase)(=C=C=), which only possesses double bonds\cite{pigulski2019reactivity}. Among these two phases, polyyne phase possessing  Peierls’ distortion is the most stable.  Carbyne-based materials have potential for use in a variety of applications, including biomedical probes, nonlinear optics, luminescence, friction-resistant coatings, and miniature electronic circuits\cite{yang2022synthesis}. Their rigidity and high conductivity set them apart from other organic chains, making them one of the best possible nanostructured materials for nanodevice applications\cite{cataldo2005polyynes}. Intriguing electronic and transport properties have been reported in carbyne. Prediction of spin-polarized electronic transport in carbon chains connected to graphene nanoribbons shows the possibility of using them in spintronics applications\cite{zanolli2010quantum}. Room temperature superconductivity  has been predicted theoretically in carbyne chain\cite{Little1964}. Enhancement in electrical conductivity has been shown for  CNT enclosed carbyne at a critical radius of the enclosed CNT\cite{Berdiyorov2020}. Today, sp-carbon nanostructures and molecules can often be synthesized in stable forms and then explored\cite{casari2016carbon}.  Nanocrystals of carbyne have been studied theoretically\cite{Belenkov2008} and synthesised experimentally\cite{Yang2021}. There has been experimental demonstration of single and multiple polyyne chains of finite length\cite{Casillas2014}. 
From a materials engineering perspective, tuning electronic characteristics and semiconductor-metal transitions  in low dimensional material is critical. Introduction of structural deformation, such as the use of strain and defect engineering\cite{tiwari2002strain}, control of layer number\cite{bafekry2020first}, hetero-structuring, chemical doping\cite{lei2018direct}, alloying and/or alternating compositions\cite{ning2017bandgap} have been used to achieve such transitions and/or tuning of the bandgap in bulk and two dimensional (2D) layered materials. Similar strategies such as functionalization\cite{Milani2017}, doping\cite{Tanaka2018} and strain\cite{Ming2014},  have been applied on 1D carbon chains too to tune its electronic properties. Artyukhov\cite{artyukhov2014mechanically} et. al. predicted that the Peierls' transition from symmetric cumulene to broken-symmetry polyyne structure is accelerated as the material is stretched. A further analogous research\cite{cretu2013electrical} demonstrates that the conductivity decreases considerably with increasing length of a semiconducting chain in the presence of strain. \textcolor{black}{ Recently, a system of two interacting polyyne chain of infinite length has been proposed theoretically\cite{basu2022structural} and band gap tuning and semiconductor to metal transition has been reported under strain and sliding.} However, rather than implementing adjustments to the fundamental structure of materials, unique extraneous methods ought to be employed in applications for cutting-edge electronics to tune the band gap. Applying external electric field and magnetic fields are such methods and have been applied on bulk and 2D materials\cite{wang1994magnetic,huang2016electric} to tune their properties and have been extended to 1D materials as well. Dos Santos\cite{dos2017electric} et al. suggested hybrid molecular systems with small carbon atomic chains joined by flakes resembling graphene, which demonstrated a semiconductor to metal transition when an electric field was applied. Similar investigations\cite{zhang2009external} were also conducted on Silicon nanowires, finding that the band gap drops as the electric field increases as a result of sharp decrease in conduction band maximum.
\textcolor{black}{
Such studies are very important for polyyne chains for their potential applications in devices. So far, tuning of electronic  properties  of one and two polyyne chain under an external transverse electric field has not been reported yet. Also, to use them as interconnects in nanoelectronic circuits, the stability of these carbon chain systems, especially the recently predicted two chain system  need to be investigated thoroughly under external electric field. } 



\textcolor{black}{In this study, for the first time we show} semiconductor to metal transition in single and two polyyne chains under the application of transverse electric field. 
\textcolor{black}{We have also endeavored to investigate spin asymmetry in the density of states after the metallic transition, which indicates the emergence of magnetic properties in the material.} In addition, dynamical and thermal stability investigations \textcolor{black}{ought to be} carried out to predict their suitability in nanodevice applications. We \textcolor{black}{ also aim to conduct an examination} of the vibrational properties and calculate the active Raman modes for one and two chains, offering a non-destructive tool to measure the transition  experimentally. \textcolor{black}{This study of achieving band gap tuning and electronic phase transition in one and two polyyne chains using a non-destructive and straightforward method is particularly advantageous due to the ease in seamless integration of such systems into electronic circuits. In addition to bandgap tuning, we show that the application of external electric field  introduces spin asymmetry in the density of states for both the systems studied here near the fermi level indicating spin dependent electronic properties, which will be beneficial for spintronic applications. }

\section{Computational Details}
The calculations were performed using \textit{ab initio} density functional theory (DFT) in conjunction with all-electron projector augmented wave potentials (PAW)\cite{bl1994hl,kresse1999phys} and the Perdew-Burke-Ernzerhof\cite{perdew1996generalized} generalized gradient approximation (PBE-GGA) to the
electronic exchange and correlation, as implemented in the Vienna \textit{ab initio} simulation package (VASP)\cite {Kresse1993,Kresse1996}.
\textcolor{black}{VASP is chosen for its proficiency in handling periodic systems using a Plane Wave basis set, known for its accuracy in electronic structure calculations.  The hybrid functional Becke, 3-parameter, Lee–Yang–Parr (B3LYP)\cite{stephens1994ab} is selected for structure optimization, aligning closely with bond length and band gap values from existing literature. This optimized structure was used in all the subsequent calculations using PBE to expedite the process, focusing on trends in band gap reduction. Cross-checking with B3LYP is conducted later to ensure the metallic transition.} 
A well-converged Monkhorst-Pack k-point set 11 × 1 × 1 was used for the calculation and the conjugate gradient scheme was employed to optimize the geometries until the forces on every atom were $\leq$0.001 eV/\si{\angstrom}. Sufficient vacuum was applied along the two directions perpendicular to the chain to avoid any interactions between periodic images. The optB88-vdW\cite{klimevs2009chemical} functional was used to accommodate van der Waals interactions. In addition to the semilocal exchange-correlation energy, a non-local correlation energy was incorporated. It was necessary to make adjustments to get contributions from the atom's surrounding spheres that were reasonably accurate because van der Waals density functionals (vdW-DF) typically produce less spherical densities than standard GGA functionals. Electronic properties and charge density difference plots were generated using sumo\cite{ganose2018sumo} and \textcolor{black}{ Visualization for electronic and structural analysis (VESTA)}\cite{momma2011vesta} respectively.

The phonopy package\cite{phonopy} was used in combination with VASP to produce phonon frequencies. A well-converged Monkhorst-Pack k-point set 5 × 1 × 1 was used and a 10 x 1 x 1 supercell was taken for the calculations. To obtain the force constants (FCs) in phonopy, density functional perturbation theory (DFPT)\cite{baroni1987green,hedin1965new} was employed and it was post-processed to find the Raman active modes. The off-resonant Raman activity was found by applying the method confirmed by Porezag and Pederson\cite{porezag1996infrared} as executed in the vasp$\_$raman.py script\cite{fonari2013raman}.  The thermal stability calculations were carried out around 300K via \textit{Ab Initio} Molecular Dynamics (AIMD) simulations\cite{car1985unified} employing VASP code. We used the canonical ensemble in our \textcolor{black}{molecular Dynamics (MD)} simulations for atomic structure thermal equilibration. The VASP package implemented the Noose-Hoover thermostat in this ensemble\cite{evans1985nose}. All the calculations were carried out for 3000 time steps where each time step is 0.25 fs. Kinetic energy cutoff was set to be 600 eV. The electric field value used for dynamical and thermal stability calculations is 1.2 (1.23) eV/\si{\angstrom} for single (two) polyyne chain, which is the transitional electric field value.

\section{Results and Discussion}
\subsection{Structural and electronic properties}
\subsection*{Single polyyne  chain}

\begin{figure}[htb!]
	\centering
	\includegraphics[width=\columnwidth]{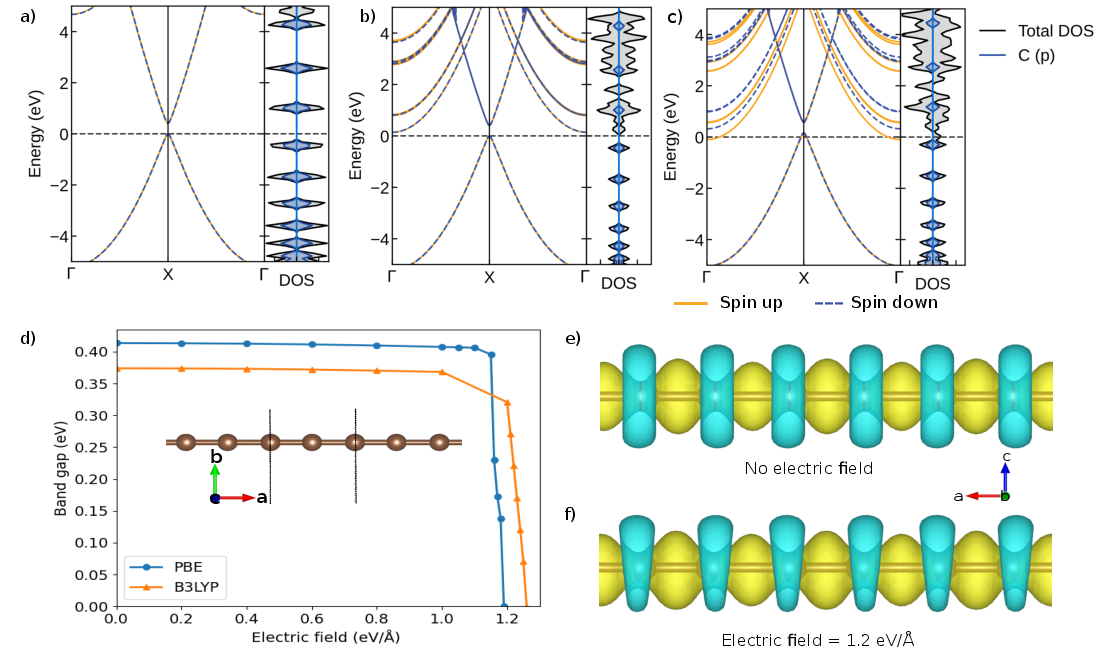}
	\caption{ Band structure of single polyyne chain at applied electric field values of a) 0, b) 1.18 and c) 1.2 eV/\AA. d) Band gap vs electric field using  PBE (with B3LYP geometry) and B3LYP functional (Inset - single polyyne chain unit cell). Charge density difference plots at e) 0 and f) 1.2 eV/\AA. Yellow and cyan represent charge accumulation and charge depletion, respectively. }
	\label{fig1}
\end{figure}

Polyyne is a linear atomic chain of carbon with alternate single and double bond between the carbon atoms. Due to the  Peierls’ distortion, the successive bond lengths are not equal giving rise to a non zero bond length alternation (BLA), defined as the difference ($\delta r$) between the long and the short bond length between the carbon atoms.  Lattice parameter optimisation as well as optimisation of BLA were carried out using PBE-GGA functional and hybrid functional  B3LYP\cite{stephens1994ab}. The unit cells used for the calculations are shown in Fig. \ref{fig1}(d) inset for the single chain. The direction of the chain was considered periodic along the x-axis and sufficient vacuum were used along the directions perpendicular to the
chain length (i.e., along y- and z-axis) to avoid any interaction between the periodic images. The structural optimization results are tabulated in Table \ref{table1}. \textcolor{black}{Results are comparable to some of the previously reported values for infinite polyyne where optimal bond lengths were identified at 1.257$\si{\angstrom}$ and 1.291$\si{\angstrom}$ with  $\delta$r=0.034$\si{\angstrom}$ using Local-density approximations (LDA) functional
\cite{bla2005}, whereas in our B3LYP calculations, the respective values are 1.251$\si{\angstrom}$, 1.295$\si{\angstrom}$, and $\delta$r=0.046$\si{\angstrom}$.} Since PBE  can not describe Peierls’ distortion, the structure relaxed into the metallic cumulene phase having a zero BLA.  Based on available experimental data, the B3LYP functional is known to produce the best predictions for both geometry and band gap of polyyne\cite{yang2006bond}. We have used \textcolor{black} {PBE-GGA functional with} structural parameters of B3LYP optimization in subsequent computations \textcolor{black}{unless mentioned otherwise}.\\

\begin{table}[h]
	\begin{tabular}{|l|c|l|c| }
        \hline
		Functional & Lattice parameter (\si{\angstrom}) &  $\delta$r (\si{\angstrom}) & Band Gap (eV) \\  \hline
		PBE        & 2.560                     & 0.000  & 0 (cumulene)\\ \hline
		B3LYP      & 2.546                      & 0.046 & 0.3739\\ \hline
	\end{tabular}
	\caption{Optimised structural parameters for single polyyne chain}
	\label{table1}
\end{table}

 To understand the electronic properties of single polyyne chain, spin polarized projected density of states  (PDOS) and band structure were calculated using PBE functional. The band structure is  plotted along the high symmetry lines in the Brillouin zone as shown in Fig. \ref{fig1}(a). The plot shows that the single polyyne chain is a semiconductor with a direct band gap of 0.414 eV at the X point. The low energy band dispersion is parabolic near the X point. The projected band structure (not shown here) reveals that the low energy bands originate from carbon $p_y$ and  $p_z$ orbitals. Fig. \ref{fig1}(a) shows symmetric nature of the DOS for both spin up and down. 
  To investigate the effect of external electric field on the electronic properties of single polyyne chain, an electric field is applied by introducing an artificial dipole sheet\cite{neugebauer1992adsorbate} in the middle of the vacuum whose center lies in the center of mass of our unit cell as implemented in VASP. Vacuum region is chosen in such a way as to restrict field emission into the vacuum\cite{feibelman2001surface}. Spin-polarised band structures of single polyyne chain were calculated under the application of transverse electric field of varying magnitudes. As observed in Fig. \ref{fig1}(a-c), spin degeneracy of bands gets lifted and splitting of bands was observed  under applied electric field for the higher energy bands. Spin up (down) bands moves closer (away) to (from) the Fermi level at the Gamma ($\Gamma$) point. The overall band gap remains almost constant until the applied electric field was (1.15 eV/\AA), when the band gap becomes indirect with the conduction band minima (CBM) at the $\Gamma$ point. With further increase in electric field, the conduction band at the $\Gamma$ point moves towards the Fermi level, causing the band gap to decrease very sharply. The single polyyne chain becomes metallic at 1.2 eV/\AA~ when the CBM (spin up) and valence band maxima (VBM) crosses the Fermi level simultaneously at the $\Gamma$ point and the X point, respectively. An asymmetry in the DOS is clearly visible for spin up and spin down states after transition in Fig. \ref{fig1}(c). Band structures and DOS plots for the intermediate electric field values are given in the Supplementary information.
  As the PBE functional underestimates band gap\cite{mori2008localization}, we performed the band structure calculations using B3LYP functional. We have observed similar trends in the band gap Vs. applied electric field plots for this hybrid functional, as shown in  Fig. \ref{fig1}(d). However, semiconductor to metal transition occurred at a slightly higher electric field (1.26 eV/\AA) for B3LYP functional.

 To understand the charge redistribution while forming the single polyyne chain from the constituent carbon atoms, charge density difference was calculated with and without external electric field (Fig. \ref{fig1}(e, f)). The charge density difference plots  were calculated by taking the difference between the charge density of the single chain and the individual charge densities of each of the two carbon atoms in the unit cell. It was observed that while electron density has depleted around the carbon atoms themselves, it has accumulated in the region between the carbon atoms.
 It is clear from the plots that the charge density was symmetric before an electric field was applied, but that it has subsequently evolved lopsided after the electric field was applied. The variation is more visible in the charge depletion regions.

 \subsection*{Two Polyyne Chains}
 
\begin{figure}[htb!]
	\centering
	\includegraphics[width=\columnwidth]{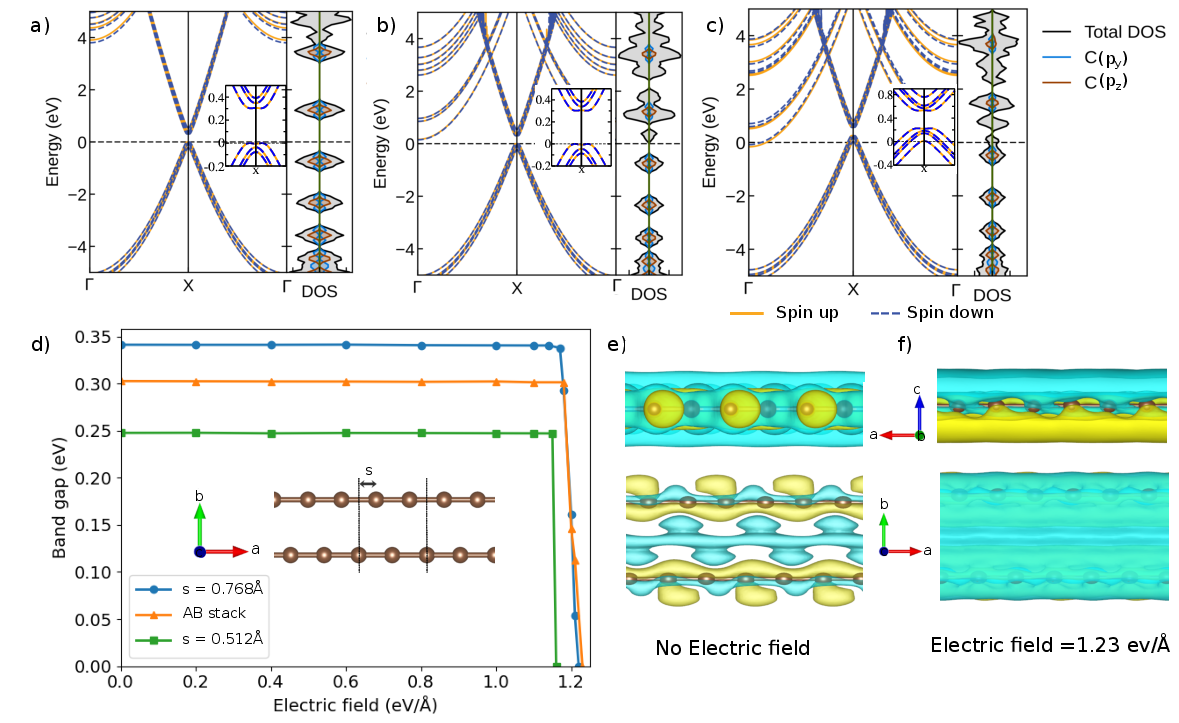}
	\caption{ Band structure of two polyyne chains at applied electric field of a) 0, b) 1.0 and c) 1.23 eV/\AA. 
 d) Band gap vs electric field plots for different sliding distances (Inset - Double chain unit cell). Charge density difference plots at e) 0 and f) 1.23 eV/\AA. Yellow and cyan represent charge accumulation and charge depletion, respectively.}
	\label{fig1a}
\end{figure}

 To understand the effect of interchain interaction on the electronic properties,  we extended our study to a system of two polyyne chains held together by weak van der Waals forces (Fig. \ref{fig1a}). We performed structural optimization to obtain interchain distance of the two polyyne chains for AB stacking first. The relaxed interchain distance was calculated to be 3.8, 3.6 and 3.6 \AA~ for PBE\cite{perdew1996generalized}+vdW, B3LYP\cite{stephens1994ab} and  LDA\cite{ceperley1980ground} functionals, respectively. We performed calculations for a few other stacking obtained by sliding one of the chains from AA stacking by an amount $s$, the sliding parameter. The pictorial definition of the stacking and the sliding parameter is shown in Fig. \ref{fig1a} (d) inset. No drastic changes in the interchain distance with slidings were observed. For the same reason as in single chain, B3LYP lattice parameters were chosen for further calculations.\\ 

   The band structure of two polyyne chains is calculated by applying transverse electric field along the direction perpendicular to the plane containing both carbon chains i.e., along the ``c" axis. The band structure of two polyyne chains close to the Fermi level as well as its band gap is very sensitive to the stacking pattern. The inset of Fig. \ref{fig1a} (a) shows a Mexican hat pattern in the energy dispersion for AB stacking. A semiconductor to metal transition was observed under the application of external electric field. With the increase in electric field, the higher energy bands in the conduction band move towards the Fermi level keeping the VBM and CBM unchanged as shown in Fig. \ref{fig1a} (b) (refer to Supplementary information for intermediate band structures). Similar to the single chain, the band gap becomes indirect and a sharp semiconductor to metal transition occur at 1.23 eV/\AA, when the VBM and CBM crosses the Fermi level (Fig. \ref{fig1a} (c)). A plot of band gap as a function of applied electric field is shown in Fig. \ref{fig1a} (d). The plot shows a constant band gap initially and then a sharp transition to the metallic state under the application of electric field.  The spin polarized DOS plots show no spin asymmetry with electric field except after metallic transition, where a spin asymmetry was observed in the conduction band. However unlike single chain, for two chains, there are finite density of states at the Fermi level for both the spins.  Similar studies were performed for two more stackings corresponding to the sliding of 0.512 \AA~ and 0.768 \AA~ and are plotted in Fig. \ref{fig1a} (d). The trends in electronic properties were similar to that of AB stacking  for these stackings as well.

 Charge density difference plots were calculated before and after the application of electric field in order to visualise the charge transfer and redistribution of charges in the two chain system. For this calculation, the charge densities of the individual chains were subtracted from the charge density of the two chain structure. Fig. \ref{fig1a} (e) and (f)  shows the charge density difference plots for the two polyyne chains system without and with electric field, respectively. The plots before the application of electric field were symmetric. Interchain interaction is clearly visible in Fig. \ref{fig1a} (e), where charge is depleted from the interchain distance and is  accumulated on the chain. After the application of electric field, the electrons accumulated in the direction opposite to the direction of applied electric field, showing charge polarity (Fig. \ref{fig1a} (f)).
 
 \subsection{Structural and thermodynamical stability and Raman active modes}
 
 From real application point of view and in order to guarantee the integrity of single and two polyyne chains  under the application of external electric field, structural stability  is very important. The methodologies, that we have utilized here to ensure structural stability of the polyyne chains under external electric field includes; formation energy calculation, phonon spectra calculation and AIMD simulation. 
 
 The formation energy for the single chain with and without external electric field was calculated as per the following equation.
 \begin{equation}
 	E_{f} = E_{tot} - E_{C1} - E_{C2}
 \end{equation}
 where, the ground state's total energy is denoted by E$_{tot}$ and E$_{C1}$ and E$_{C2}$ refers to the independent ground state energies of the two carbon atoms in the structure. The formation energies of the single polyyne chain at electric field of 0, 1.1 and 1.2 eV/\AA~ have been listed in Table. \ref{Ef}. \textcolor{black}{Our formation energy value for polyyne chain without external electric field is of the same order as reported in Ref.\cite{Li2018}.} 

 \begin{table}[h!]
 	\begin{tabular}{|c|c|c|c|}
    \hline
   \multicolumn{2}{|c|}{Single polyyne chain } &  \multicolumn{2}{c|}{Two polyyne chains} \\ \hline
 		Electric field (eV/\AA) &   E$_f$ (eV)  & Electric field (eV/\AA) &  E$_f$ (eV) \\ \hline
 		0                                     & -11.86 & 0  &  -0.047  
 		\\ \hline
 		1.1                                     & -11.79 & 0.9  & -0.036 
            \\ \hline
 		1.2                                       & -11.77 & 1.23 &  -0.021
 		\\ \hline
 	\end{tabular}
 	\caption{Formation energy values for single and two polyyne chains with electric field}
 	\label{Ef}
\end{table}
  The formation energy gives a measure of the structural stability, i.e., against any molecular desorption from the polyyne structure. It is evident from Table. \ref{Ef}  that the formation energy remained negative both before and after the metallic transition. From this result, it may be inferred that the structure is intact after the application of electric field. Repeated calculations for slightly higher electric field strengths also ended up in negative formation energies, indicating that the structure is stable and can be realized in experiments. It should be noted, however, that as the strength of the electric field increases, the values are showing a trend toward being less negative.\\
\begin{figure}[htb!]
	\centering
	\includegraphics[width=\columnwidth]{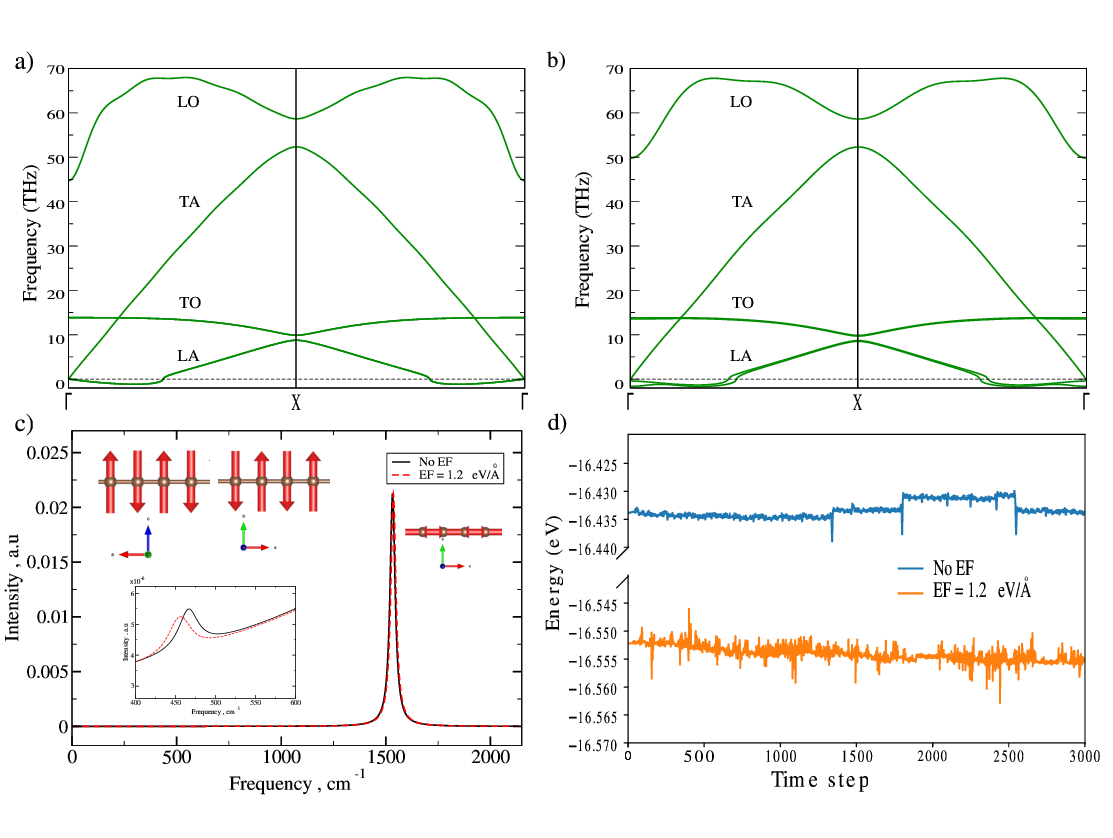}
	\caption{ Phonon dispersion spectra of single polyyne chain at a) zero electric field and b) after metallic transition at electric field = 1.2 eV/\si{\angstrom}. c) Off resonant Raman activity spectra for zero electric field and after metallic transition. The Raman active vibrations are shown in the inset.  d) Thermal stability at 300K for zero electric field and after metallic transition.}
	\label{fig2}
\end{figure}
  To ensure dynamical stability of the structure, phonon dispersion spectra, which investigates the lattice vibrational modes was calculated with and without external electric field.  From the phonon band structures shown in Fig. \ref{fig2} (a) and (b) for electric fields of zero and 1.2 eV/\AA, respectively, the structure can be regarded as dynamically stable despite the presence of some negative frequencies close to the $\Gamma$ point since those negative frequencies can be treated as computational noise. Small inaccuracies in the forces estimated by VASP, which are used to compute second derivatives numerically, are causing the imaginary frequencies that we calculated using VASP and Phonopy\cite{low2021thermodynamics}. There are total 6 modes in the phonon spectra: two doubly degenerate longitudinal (LA) and transverse acoustic (TA) modes with zero frequency at the $\Gamma$ point, two optical modes; one transverse (TO) and one longitudinal (LO) mode with frequencies of 14 and 45 THz, respectively. Compared to cumulene, polyyne phonon dispersion shows a gap between the LA and LO branches at the X point (the zone boundary  point in the Brillouin zone) attributed to Peierls' distortion. The frequency of the LO mode at the $\Gamma$ point increases from 45 THz at zero electric field to 50 THz at 1.2 eV/\AA. Slight splitting of bands were observed in LA and TO modes after the application of the transverse electric field, which  lifted their degeneracies. Further post-processing revealed that out of these modes, 3 modes are Raman active. The pictorial representation of these modes are shown in Fig. \ref{fig2} (c).  Mode 1, corresponding to the LO mode had the highest Raman activity and is the origin of the sharp peak at 1532 cm$^{-1}$ in the polyyne Raman spectra as shown in the plot of Fig. \ref{fig2} (c). This is the characteristic peak of sp-hybridized carbons chain experimentally observed in the spectral region of 1800–2300 cm$^{-1}$\cite{expRaman,expRaman1,expRaman2}. A slight shift in the peak position and increase in intensity was observed after transition, showing electro-optic effect. The other two Raman active modes, Mode 2 and Mode 3 (related to TO branch) are degenerate modes giving a low intensity peak at 466 cm$^{-1}$ shown in the inset of Fig. \ref{fig2} (c).

The thermal stability analysis was performed using AIMD simulations at 300K. Total energy was mapped against time steps upto 750 fs as shown in Fig \ref{fig2}(d) for electric field of 0 and 1.2 eV/\AA, respectively. No major abrupt change in total energy Vs. time steps was observed for both with and without electric fields. The thermal fluctuations becomes minimum after 750 fs, indicating that the structure is thermally stable at room temperature. It is evident that once the electric field was applied, the structure became even more stable; i.e. the magnitude of total energy increased from -16.43 eV to -16.55 eV.

 We extended our study on stability analysis to the two chain system. It is important to know if the system can be synthesized experimentally and will be stable under the application of electric field. We calculated the formation energy of the two polyyne chains as follows,
 
 \begin{equation}
 	E_{f} = E_{tot} - E_{Chain1} - E_{Chain2}
 \end{equation}

 where, E$_{tot}$  refers to the ground state energy of the two polyyne chains structure and E$_{Chain1}$ and E$_{Chain2}$ denotes ground state energies of the two isolated chains, respectively. The formation energy is listed in Table. \ref{Ef}. The formation energy was negative throughout the range of the applied electric field, indicating that it is energetically stable during the semiconductor to metal transition.  Decrease in magnitude of the formation energy  was observed with the increase in electric field, similar to the single chain. \\

 \begin{figure}[htb!]
	\centering
	\includegraphics[width=\columnwidth]{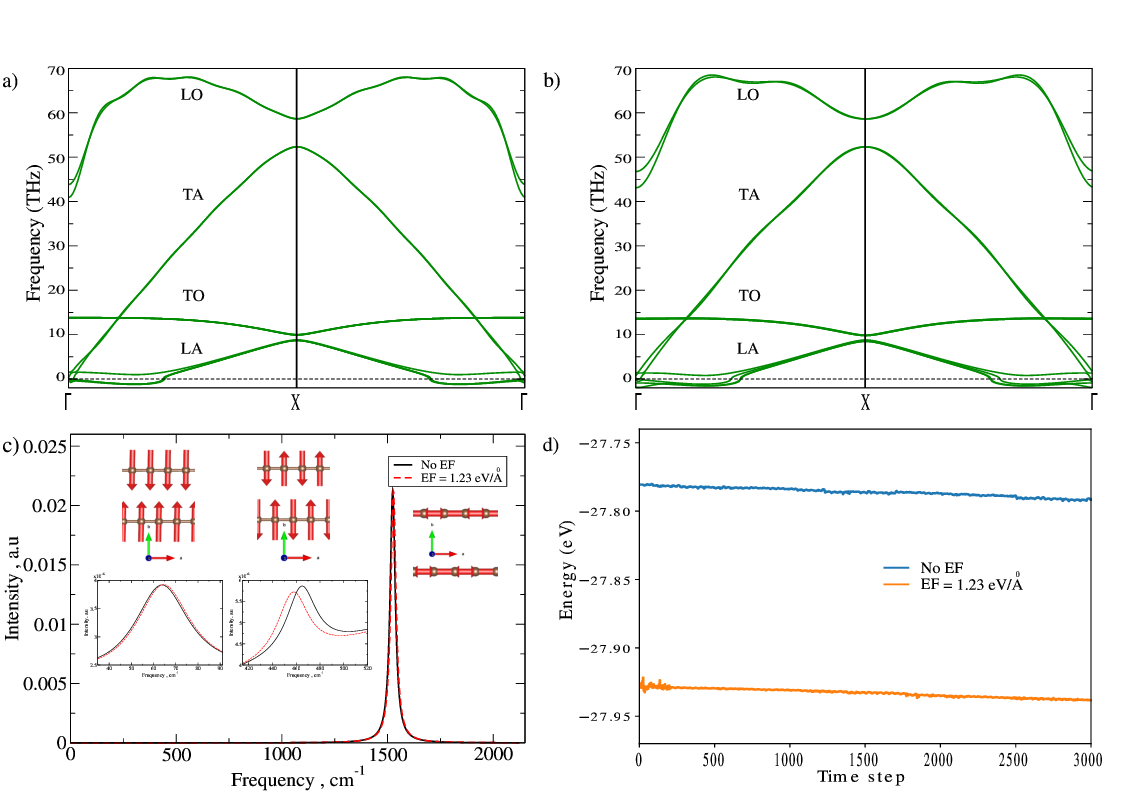}
	\caption{ Phonon dispersion spectra of two polyyne chains at a) zero electric field and b) after metallic transition at electric field = 1.23 eV/\si{\angstrom}. c) Off resonant Raman activity spectra for zero electric field and after metallic transition. The Raman active vibrations are shown in the inset.  d) Thermal stability at 300K for zero electric field and after metallic transition.}
	\label{fig2a}
\end{figure}
 
 The phonon band structure was calculated  and plotted for zero electric field and after metallic transition in Fig. \ref{fig2a} (a) and (b), respectively. The phonon band structure of two chain  resembles very much to that of the single chain. Splitting of branches was observed near the $\Gamma$ point except for the TO mode. The splitting increased slightly with increase in electric field. The frequencies of the LO modes increased slightly with electric field. Some negative frequencies were observed around the $\Gamma$ point which can be considered as computational noise. As the negative frequencies are of the order of one THz, the structure can therefore be considered dynamically stable. Further post-processing revealed that out of these, 3 modes were Raman active and Mode 1, the characteristic peak of polyyne chain had the highest activity.  The corresponding representations and Raman spectra at no electric field and after transition are shown in Fig \ref{fig2a} (c).  The peak at 1524 cm$^{-1}$ corresponds to Mode 1, a LO mode, whereas Modes 2 corresponding to peak at 464 cm$^{-1}$ and 3 corresponding to peak at 64 cm$^{-1}$ are TO and TA modes, respectively. Total energy was mapped against time steps to examine the material's thermal stability at room temperature. No abrupt change in total energy Vs. time steps was observed for both with and without electric field, according to FIG \ref{fig2a}(d). For the plot after metallic transition, large fluctuations were observed initially which reduced after 750 fs. In a manner analogous to the single chain configuration, the application of an electric field resulted in an increase in magnitude of the total energy from -27.89 eV to -28.04 eV. The total energy varies for both the plots, however, the variation is quite small compared to the formation energy of the two polyyne chains. Thus we can infer that the structure has good thermal stability with and without external electric field. Thus, from the stability analysis, we can predict that  the two polyyne chains system  can be synthesized experimentally and it will be thermally and dynamically stable during a semiconductor to metal transition under the application of an external electric field. 
  \section{\label{sec:level4}Conclusion}
 In conclusion, we show a sharp semiconductor to metal transition  in  single and two polyyne chains under the application of a transverse electric field. Both single and two chain system show quite similar response in band dispersion to the applied electric field, i. e., the high energy bands migrate towards the Fermi level and VBM and CBM cross the Fermi level simultaneously at metallic transition. A distinct spin asymmetry was observed in the density of states in single and two polyyne chains after transition. 
 \textcolor{black}{This phenomenon signifies the prevalence of electrons with specific spin orientations. Such spin-polarized states have the capability to engender spin-polarized currents. This property holds promise for the development of spintronic devices, including spin valves, and exhibits potential utility in spin injection and detection applications.} Both the systems show sharp transitions, with two chain having sharper than the single chain, in the band gap at the transition electric field. \textcolor{black}{This observation indicates the potential utility of these entities as switches and in various other applications within the field of nanoelectronics, because, the fundamental aspect of digital electronics, encompassing logic gates and microprocessors, relies on the resistive switching phenomenon, wherein transitions occur between conducting and non-conducting states.} Both the structures were found stable throughout the transition and thus can be realized experimentally. Raman spectra show characteristic peak of polyyne in both the single and two chains and additional peak with low intensity in two chain system.  Shift in frequencies of some of the vibrational modes under the application of electric field offers a non-invasive characterization technique to measure these transitions in experiments.

 \textcolor{black}{
 The presented results give an indication for experimental realisation of such transitions and the applicability of these chain systems in next generation nanodevices. However, there will be slight deviations in real experiments as our results are limited by the accurate geometry and band gap, both being underestimated by the computational methodology (DFT) used in this work.  In practical scenarios, where finite chains or individual molecules are involved, it may necessitate a transition from Plane Wave to Gaussian or Localized basis sets. This could necessitate sensitivity analyses on basis sets, referencing higher-level theoretical/experimental outcomes. Additionally, scalability concerns emerge with increasing system size, demanding a more accurate refinement of the functional for modeling electron exchange and correlation effects. Although polyyne in its pure form is not known to pose any health hazard, some solvents used in its synthesis can be toxic. Moreover, solvents can significantly impact the electronic structure, geometry, and energetics of molecules, introducing interactions like solute-solvent interactions and solvent polarization effects, which can be a future study. }

\subsection*{CRediT authorship contribution statement}

\textbf{Karthik H J (First Author)}: Conceptualization, Data Curation, Methodology, Investigation, Formal Analysis, Writing - Original Draft, Review \& Editing. \textbf{Sarga P K}: Investigation, Methodology, Data curation. \textbf{Swastibrata Bhattacharyya (Corresponding Author)}: Conceptualization, Methodology, Validation, Formal Analysis,  Resources, Supervision,  Writing - Review \& Editing.

\begin{acknowledgments}
SB would like to acknowledge SERB, Govt. of India (SRG/2020/000562 and CRG/2020/000434) and BITS Pilani K. K. Birla Goa Campus, India (GOA/ACG/2019-20/NOV/08 and C1/23/211) for the financial support. SPK acknowledges SERB, Govt. of India (SRG/2020/000562) for financial support. All the authors acknowledge the High-Performance Computing Facility (HPC), BITS Pilani KK Birla Goa Campus.
\end{acknowledgments}

\subsection*{Data availability}
Data will be made available on request.
\bibliography{KHJPolyyne.bib}

\end{document}